\documentclass[5p,sort&compress,times,letterpaper]{elsarticle}
\usepackage{amsmath,amssymb}

\usepackage{bm}
\usepackage{graphicx}
\usepackage{color}

\usepackage{epsfig}
\usepackage{graphicx}
\usepackage{dcolumn}
\usepackage{bm}
\usepackage{multirow}
\usepackage{array}
\usepackage{slashed}
\usepackage{amsmath}
\usepackage[titletoc]{appendix}
\usepackage{color}
\usepackage{diagbox}

\begin{document}

\title{Presupernova neutrino signals as potential probes of neutrino mass hierarchy}

\author[GSI]{Gang Guo}
\author[UMN,TDLI]{Yong-Zhong Qian}
\author[MU,TDLI]{Alexander Heger}
\address[GSI]{GSI Helmholtzzentrum f{\"u}r Schwerionenforschung, Planckstra$\ss$e 1, 64291 Darmstadt, Germany}
\address[UMN]{School of Physics and Astronomy, University of Minnesota, Minneapolis, MN 55455, USA}
\address[MU]{Monash Centre for Astrophysics, School of Physics and Astronomy, Monash University, VIC 3800, Australia}
\address[TDLI]{Tsung-Dao Lee Institute, Shanghai 200240, China}

\date{\today}

\begin{abstract}
We assess the potential of using presupernova neutrino signals at the Jiangmen Underground Neutrino Observatory (JUNO)
to probe the yet-unknown neutrino mass hierarchy. Using models for stars of 12, 15, 20, and $25\,M_\odot$,
we find that if the $\bar\nu_e$ signals from such a star can be predicted precisely and the star is within 
$\approx 440$--880~pc, the number of $\bar\nu_e+p\to n+e^+$ events detected within one day of its explosion allows to 
determine the hierarchy at the $\gtrsim 95\%$ confidence level.     
For determination at this level using such signals from Betelgeuse, which is at a distance of
$\approx222$~pc, the uncertainty in the predicted number of signals needs to be $\lesssim 14$--30\%.
In view of more realistic uncertainties, we discuss and advocate a model-independent determination using both $\nu_e$ 
and $\bar\nu_{e}$ signals from Betelgeuse. This method is feasible if the cosmogenic background for $\nu$-$e$ scattering 
events can be reduced by a factor of $\sim 2.5$--10 from the current estimate. Such reduction might be achieved by
using coincidence of the background events, the exploration of which for JUNO is highly desirable.
\end{abstract}    

\maketitle

\section{Introduction} \label{sec:intro}
Stars are profuse sources of neutrinos. For massive stars of $\gtrsim 8 M_\odot$, 
as their central temperature and density increase dramatically during later evolution stages, 
$\nu_a\bar\nu_a$ ($a=e,\mu,\tau$) pair production by photo-neutrino emission, plasmon decay, and
$e^\pm$ pair annihilation becomes the dominant mechanism of energy loss (e.g., \cite{Itoh96,guo}). 
Likewise, $\nu_e$ and $\bar\nu_e$ production by weak nuclear processes, including $e^\pm$ 
capture and $\beta^\pm$ decay, becomes more and more significant as such stars evolve. 
These neutrinos not only play essential roles in cooling the interiors of massive stars, 
but also serve as potential signatures of their evolution, which leads to 
the eventual core collapse and supernova (SN) explosion. With the next generation of 
detectors such as the Jiangmen Underground Neutrino Observatory (JUNO) \cite{juno} 
and the Deep Underground Neutrino Experiment (DUNE) \cite{dune} under construction, 
there is growing interest in detecting pre-SN neutrinos. 
Previous studies 
\cite{odrzywolek04,odrzywolek10,kato15,kamland16,yoshida16,kato17,patton1,patton2} 
showed that it is plausible to detect the pre-SN $\bar\nu_e$ from a star 
within a few kpc a few days before its explosion, thereby providing an advance warning. 
A promising candidate is Betelgeuse with an estimated mass of $20_{-3}^{+5}\,M_\odot$
\cite{progenitor} and at a distance of $222^{+48}_{-34}$~pc \cite{distance}.

In this paper we focus on the possibility of using pre-SN neutrinos to determine the 
yet-unknown neutrino mass hierarchy ($\nu$MH). As these neutrinos propagate through the stellar 
interior, they undergo flavor transformation due to the Mikheyev-Smirnov-Wolfenstein (MSW) 
effect \cite{msw}. This effect depends on the electron number density profile of the star  
and the vacuum neutrino mixing parameters, especially on whether the $\nu$MH
is normal (NH) or inverted (IH) \cite{starosc}. Because the survival probability of $\bar\nu_e$
for the NH is much higher than that for the IH, the rate of 
$\bar\nu_e+p\to n+e^+$ (inverse $\beta$-decay, IBD) events in a detector is correspondingly 
higher for the NH \cite{kato15,kamland16,kato17}. Unaware of any detailed analysis, here we 
quantitatively assess the potential of using pre-SN neutrino signals as probes of the $\nu$MH.

Based on the typical energies and fluxes of pre-SN neutrinos, we focus on JUNO as the
detector, whose best capability is to detect $\bar\nu_e$ above $\approx1.8$~MeV through IBD.
The key input to determine the $\nu$MH from pre-SN $\bar\nu_e$ signals is
the theoretical model for the stellar source. We adopt representative
models \cite{alex} for stars of 12, 15, 20, and $25\,M_\odot$. For each
model, we determine the limiting distance within which the NH or IH can be distinguished
assuming that the predicted number of IBD signals is precise. We further estimate the maximum
uncertainty permitted in the prediction so that such signals from Betelgeuse can be used to 
determine the $\nu$MH. In view of realistic uncertainties, we finally discuss a 
model-independent determination using both IBD and $\nu$-$e$ scattering (ES) events
at JUNO.

\section{Analyses with IBD events only}  
\label{sec:ibd}
\begin{figure*}[htbp] 
\centering
\includegraphics[width=8.cm]{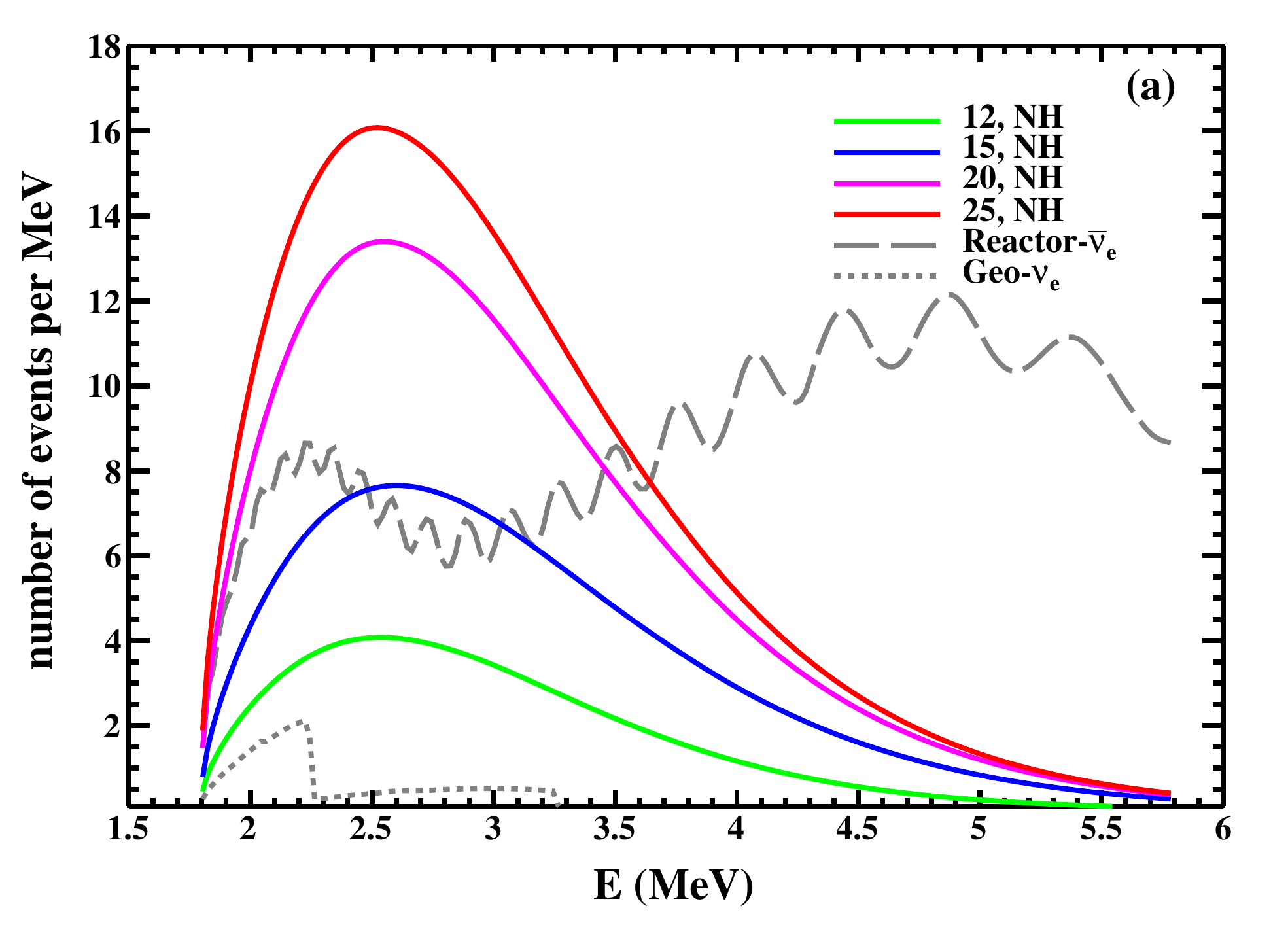}     
\includegraphics[width=8.cm]{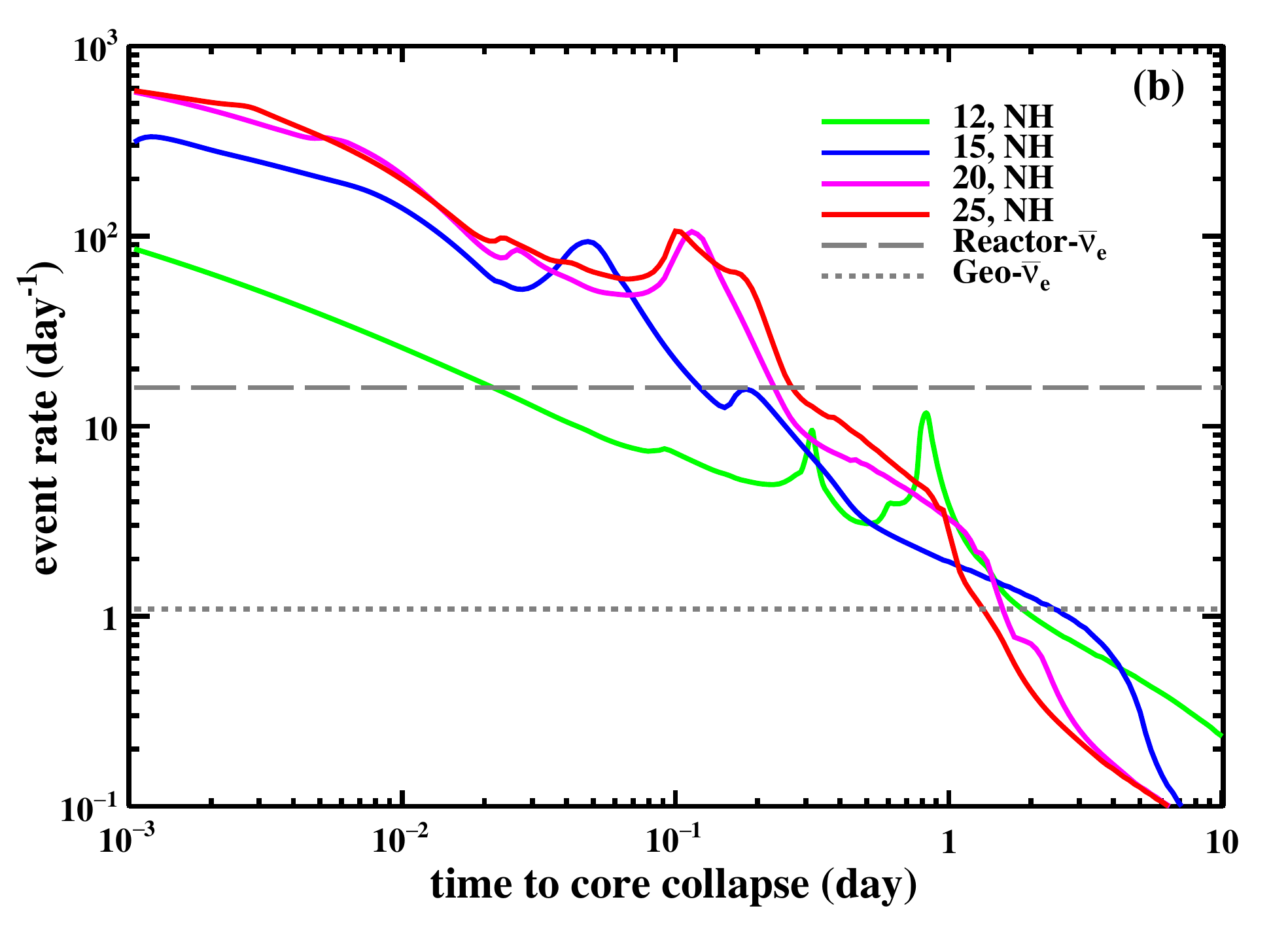}   \\  
\caption{Expected spectra (a) and time evolution of rates (b) for IBD events at JUNO 
for stars of 12, 15, 20, and $25\,M_\odot$ at $d=1$ kpc. 
The spectra are integrated over the last day before core collapse and the rates 
are for the $\bar\nu_e$ energy window of $1.8\le E\le 4$ MeV. Both these results are 
shown for the NH and should be reduced by a factor of $\approx 3.4$ for the IH.
The backgrounds from reactor $\bar\nu_e$ and geo-$\bar\nu_e$ are also shown for 
comparison.} 
\label{fig:events} 
\end{figure*}

Pre-SN $\bar\nu_e$ signals mostly occur a few days prior to core collapse and 
are predominantly produced by $e^\pm$ pair annihilation in a star.
Weak nuclear processes contribute significantly to these signals within $\sim 1$~hour 
of the core collapse \cite{odrzywolek10}, but account for 
$\lesssim 10\%$ of the total pre-SN $\bar\nu_e$ signals 
\cite{kato17,patton1,patton2}. Below we only consider the signals 
from $e^\pm$ pair annihilation.

Without neutrino oscillations, the energy-differential pre-SN $\bar\nu_a$ flux
from a star is 
\begin{align}
F^{(0)}_{\bar\nu_a}(E, t) =\frac{1}{4\pi d^2} 
\int j_{\bar\nu_a}(E, T, n_e, t)dV,  
\end{align}  
where $E$ is the $\bar\nu_a$ energy, $t$ is time, $d$ is the distance to the star,
$j_{\bar\nu_a}$ is the 
energy-differential rate of $\bar\nu_a$ production by $e^\pm$ pair annihilation per unit 
stellar volume, and $dV$ is the differential volume element. The calculation of 
$j_{\bar\nu_a}$ requires the temperature, $T$, and the net electron number density, $n_e$, 
both of which vary with the radius inside the star and with time.  

Pre-SN $\bar\nu_a$ undergo flavor transformation due to
the MSW effect \cite{msw}. Inspection of the stellar $n_e$ profiles shows that 
flavor evolution of pre-SN $\bar\nu_a$ with $1\lesssim E\lesssim 10$~MeV is
highly adiabatic. Therefore, the $F_{\bar\nu_e}(E,t)$ at JUNO is
\begin{align}
F_{\bar\nu_{e}}(E,t)=\bar p F_{\bar\nu_{e}}^{(0)}(E,t) + (1-\bar p) F_{\bar\nu_{x}}^{(0)}(E,t),
\label{eq:Fnu}
\end{align}
where $\bar\nu_x$ is equivalent to $\bar\nu_\mu$ or $\bar\nu_\tau$,
$\bar p=\cos^2\theta_{12}\cos^2\theta_{13} \approx 0.681$ for the NH, and 
$\bar p=\sin^2\theta_{13} \approx 0.022$ for the IH \cite{starosc,numix,pdg}. 
For the time window and $\bar\nu_a$ energy relevant for detection, we find that 
$F^{0}_{\bar\nu_x}(E,t)/F^{0}_{\bar\nu_e}(E,t)\approx 0.2$. Consequently,
$F_{\bar\nu_{e}}(E,t)$ for the NH is $\approx 3.4$ times higher than that
for the IH. We use detailed stellar models \cite{alex}
to calculate $F_{\bar\nu_{e}}(E,t)$.

The energy spectrum of pre-SN IBD events integrated
over a time window [$t_1, t_2$] is
\begin{align}
\frac{dN_{\rm IBD}}{dE}=N_p\int_{t_1}^{t_2}F_{\bar\nu_e}(E,t)
\sigma_{\rm IBD}(E)\epsilon(E)dt,
\end{align}
where $N_p$ is the total number of protons in JUNO (20 kton liquid scintillator 
with a proton mass fraction of $\approx12$\%), $\sigma_{\rm IBD}(E)$ is the IBD 
cross section, and $\epsilon(E)\approx0.73$ is the detection efficiency \cite{juno}. 
In Fig.~\ref{fig:events}a, we show the $dN_{\rm IBD}/dE$ over the last day prior to
the core collapse at $d=1$~kpc for four stellar models \cite{alex} and the NH.
For comparison, we also show the expected background, which is predominantly 
from the two closest reactors with negligible contributions from 
geo-$\bar\nu_e$ \cite{geo}. As shown in
Fig.~\ref{fig:events}a, pre-SN IBD spectra peak at $\sim 2.5$~MeV and 
decrease rapidly above $\sim 4$~MeV, where the reactor $\bar\nu_e$
background dominates. For all the results on the IBD signals presented below, 
we adopt the $\bar\nu_e$ energy window of $1.8\le E \le 4$~MeV, where the lower 
value corresponds to the IBD threshold. We find that this choice is close to optimal 
for analyzing these signals. Within this energy window
and over the last day prior to the core collapse at $d=1$~kpc,
we expect 6.1 (1.9), 12.0 (3.6), 20.5 (5.9) and 24.5 (7.0) IBD signals in 
JUNO for the NH (IH) using stellar models \cite{alex} of 12, 15, 20 and $25\,M_\odot$,   
respectively. For comparison, 15.7 and 1.1 events are expected from reactor $\bar\nu_e$
and geo-$\bar\nu_e$, respectively. The corresponding rates are
shown as functions of time in Fig.~\ref{fig:events}b.

We now estimate the limiting distance $d_{\rm lim}$ within which pre-SN IBD 
signals might allow a determination of the $\nu$MH. For each of our adopted stellar 
models, we calculate the predicted number, $N_{\rm IBD}$, of IBD events with 
$1.8\le E \le 4$~MeV and over the time window $[t_1,t_2]$ as a function of $d$ 
and $\Delta=t_2-t_1$, where $t_2$ always corresponds to the onset of core collapse.
We then determine how likely the cases of the NH and IH can be distinguished
considering the background, statistical fluctuations, and uncertainty in $N_{\rm IBD}$. 

We assume that the relative uncertainty $\alpha$ of  
$N_{\rm IBD}$ follows a Gaussian distribution 
$G(\alpha)\propto\exp[-\alpha^2/(2\sigma_\alpha^2)]$ normalized over
$-1\leq\alpha<\infty$, and that the expected number,
$N_b^{\rm IBD}$, of background events is well measured. Under these assumptions, 
the observed number of events, $N$, follows the distribution
\begin{align}
P(N,N_b^{\rm IBD},N_{\rm IBD},\sigma_\alpha)=\int_{-1}^\infty
\frac{G(\alpha)[N_{\rm IBD}(1+\alpha)+N_b^{\rm IBD}]^N}
{N!\exp[N_{\rm IBD}(1+\alpha)+N_b^{\rm IBD}]}d\alpha.
\label{eq:poisson}
\end{align}
For a fixed set of $N_b^{\rm IBD}$, $N_{\rm IBD}^{\rm NH}$, $N_{\rm IBD}^{\rm IH}$, 
and $\sigma_\alpha$, the distributions $P(N,N_b^{\rm IBD},N_{\rm IBD}^{\rm NH},\sigma_\alpha)$ 
and $P(N,N_b^{\rm IBD},N_{\rm IBD}^{\rm IH},\sigma_\alpha)$ cross at $N=N_0$,
where $N_{\rm IBD}^{\rm NH}$ and $N_{\rm IBD}^{\rm IH}$ are the predicted numbers of
signals for the NH and IH, respectively.
If the NH is true, then the probability of observing more than $N_0$ events is
\begin{align}
P_{\rm NH}^{\rm IBD}= \sum_{N=N_0+1}^\infty P(N,N_b^{\rm IBD},N_{\rm IBD}^{\rm NH},\sigma_\alpha).
\label{eq:pnh}
\end{align} 
Given that $N_{\rm IBD}^{\rm NH}\approx 3.4N_{\rm IBD}^{\rm IH}$, the above
outcome can be distinguished from the case of the IH at a confidence level (CL) of
\begin{align}
P_{\rm IH}^{\rm IBD}= \sum_{N=0}^{N_0}P(N,N_b^{\rm IBD},N_{\rm IBD}^{\rm IH},\sigma_\alpha). 
\label{eq:pih}
\end{align}
Consequently, we have a probability of $P_{\rm NH}^{\rm IBD}$ to exclude the IH at a CL of 
$P_{\rm IH}^{\rm IBD}$ if the NH is true. Likewise, if the IH is true, we have a probability of
$P_{\rm IH}^{\rm IBD}$ to exclude the NH at a CL of  $P_{\rm NH}^{\rm IBD}$. We take
$P_{\rm NH}^{\rm IBD}=P_{\rm IH}^{\rm IBD}=95\%$ and refer to fulfillment of this criterion as
determining the $\nu$MH at the 95\% CL.
 
\begin{figure}[htbp]  
\centering
\includegraphics[width=8.cm]{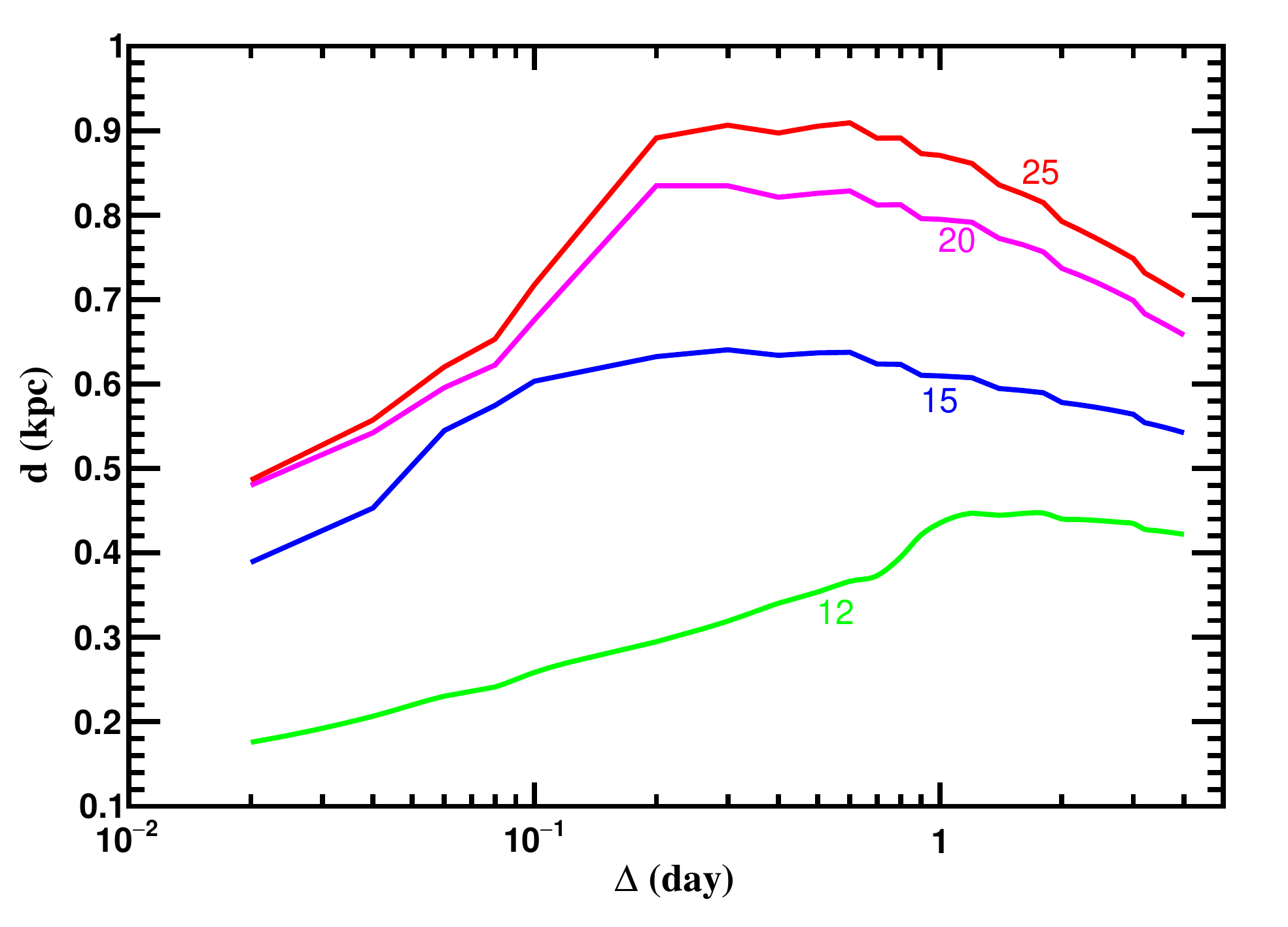}     
\caption{Combinations of $d$ and $\Delta$ for which the $\nu$MH can be determined at the 95\% CL 
in the ideal case with precisely-predicted numbers of pre-SN IBD signals from stars of 12, 15, 20,
and $25\,M_\odot$.} 
\label{fig:dis} 
\end{figure}

To precisely predict $N_{\rm IBD}$,
we must know with high accuracy the distance $d$ to the source and
its stellar model for pre-SN neutrino emission. 
Assuming that $d$ is known exactly, we consider an ideal case of
precisely predicted $N_{\rm IBD}$ by taking $\sigma_\alpha = 10\%$ 
for the uncertainty in the stellar model. 
For this case, we show in Fig.~\ref{fig:dis} combinations of $d$ and $\Delta$ for which
the $\nu$MH can be determined at the 95\% CL for each of the adopted stellar models. 
It can be seen that the largest $d$ values correspond to
$\Delta\sim1$--4, 0.1--1, 0.2--1, and 0.2--1 day for stars of
12, 15, 20, and 25 $M_\odot$, respectively. Taking $\Delta=1$~day,
we obtain $d_{\rm lim}\approx 0.44$, 0.6, 0.8, and 0.88~kpc, respectively, as the 
limiting distance within which the $\nu$MH can be determined at the $\gtrsim 95\%$ CL for 
the ideal case. We find that $\Delta=1$~day is not only optimal 
for all of our stellar models in this case, but also for $\sigma_\alpha\gg 10\%$. 
We take $\Delta=1$~day for all the analyses below.

\begin{figure}[htbp] 
\centering
\includegraphics[width=8.cm]{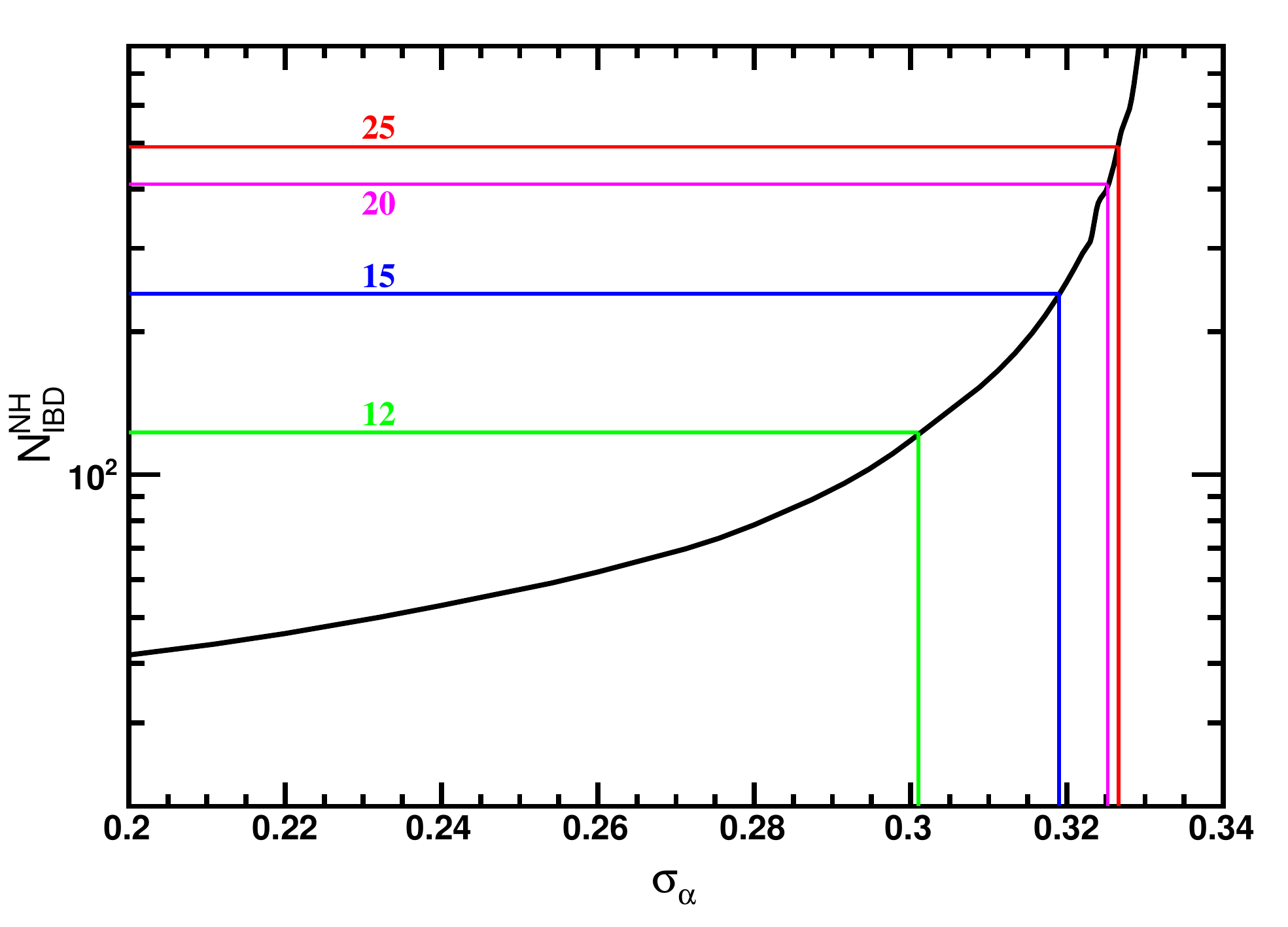}        
\caption{Combinations of $N^{\rm NH}_{\rm IBD}$ and $\sigma_\alpha$ required to determine 
the $\nu$MH at the 95\% CL. The horizontal solid lines indicate the predicted 
$N_{\rm IBD}^{\rm NH}$ over the last day before the core collapse of Betelgeuse at 
$d=222$ pc for an assumed mass of 12, 15, 20, or $25\,M_\odot$.  
}
\label{fig:mu-alpha}   
\end{figure}

For a specific source, $N_{\rm IBD}^{\rm NH}$ and $N_{\rm IBD}^{\rm IH}$ 
are related by a fixed factor and have the same relative uncertainty $\sigma_\alpha$.
Using Eqs.~(\ref{eq:poisson}), (\ref{eq:pnh}), and (\ref{eq:pih}),
we show in Fig.~\ref{fig:mu-alpha} the combinations of $N_{\rm IBD}^{\rm NH}$ 
and $\sigma_\alpha$ that are required to determine the $\nu$MH at the 95\% CL.
As an example of using this figure, we assume that one of our stellar models provides 
a good description of Betelgeuse as a potential source. 
We take $d=222$~pc and show the $N_{\rm IBD}^{\rm NH}$ predicted by
our models in Fig.~\ref{fig:mu-alpha}.
It can be seen that if one of these models fits Betelgeuse, the uncertainty in the 
predicted $N_{\rm IBD}^{\rm NH}$ is required to be $\sigma_\alpha\lesssim 30\%$
so that its pre-SN IBD signals can be used to determine the $\nu$MH at the $\gtrsim 95\%$ CL.
With the current measurement of $d=222^{+48}_{-34}$~pc for Betelgeuse
\cite{distance}, the error in $d$ already contributes $\sim 30\%$ to $\sigma_\alpha$,
which leaves little room for error in stellar models.
An uncertainty of $\sim 30\%$ in the model prediction 
is permitted, however, if a precise distance measurement, e.g.,
at the $\sim 1\%$ level becomes available.

It is unclear which of our stellar models fits Betelgeuse. This uncertainty greatly 
increases the error in predicting its pre-SN IBD signals. Consistent with the mass estimate 
of Ref.~\cite{progenitor}, we assume that our 15 and $25\,M_\odot$ models
represent the limiting cases for Betelgeuse. Under this assumption, we estimate the
restriction on $\sigma_\alpha$ so that the case of a $15\,M_\odot$ star and the NH can be
distinguished from that of a $25\,M_\odot$ star and the IH.
Using $N_{\rm IBD}^{\rm NH}$ for a $15\,M_\odot$ star in Eq.~(\ref{eq:pnh}) and
$N_{\rm IBD}^{\rm IH}$ for a $25\,M_\odot$ star in Eq.~(\ref{eq:pih}) and assuming the same
$\sigma_\alpha$ for both these numbers, we find that $\sigma_\alpha\lesssim 14\%$ is required
to distinguish the two cases at the $\gtrsim 95\%$ CL.
This requirement is unlikely to be fulfilled by stellar
models even if the distance to Betelgeuse can be measured precisely.
Clearly, a model-independent determination of the $\nu$MH is highly desirable.
Below we discuss such a determination using both the pre-SN IBD and ES events at JUNO.

\section{Model-independent analyses}
\label{sec:es}
All neutrino species contribute to the ES events.
Subsequent to flavor evolution in the stellar interior, the pre-SN
neutrino fluxes at JUNO for species other than $\bar\nu_e$ are
\begin{align}
 & F_{\bar\nu_\mu+\bar\nu_\tau}(E,t)
  =(1-\bar p)F_{\bar\nu_{e}}^{(0)}(E,t) + (1+\bar p) F_{\bar\nu_{x}}^{(0)}(E,t), \\
 & F_{\nu_{e}}(E,t)=p F_{\nu_{e}}^{(0)}(E,t) + (1-p) F_{\nu_{x}}^{(0)}(E,t), \\
 & F_{\nu_\mu+\nu_\tau}(E,t)
 =(1-p) F_{\nu_{e}}^{(0)}(E,t) + (1+p) F_{\nu_{x}}^{(0)}(E,t),   
\end{align}
where $p= \sin^2\theta_{13} \approx 0.022$ for the NH, and 
$p=\sin^2\theta_{12}\cos^2\theta_{13} \\\approx 0.291$ for the IH \cite{starosc,numix,pdg}. 
Considering recoil electrons with kinetic energy $T_{e,1}\leq T_e\leq T_{e,2}$
and assuming 100\% detection efficiency, we estimate the expected number, 
$N_{\rm ES}$, of ES events as 
\begin{align}
N_{\rm ES}= N_e\int_{t_1}^{t_2}dt\int_{E_1}^\infty dE
\int_{T_{e,1}}^{T_{e,u}}
\sum_\nu F_{\nu}(E, t) \frac{d\sigma_{\nu e}(E,T_e)}{dT_e}dT_e,
\label{eq:es}
\end{align}                       
where $N_e$ is the total number of electrons in JUNO,
$T_{e,u}={\rm min}\{T_{e,2},T_e^{\rm max}\}$,
$T_e^{\rm max}=E/[1+(2m_e/E)]$, $m_e$ is the
electron rest mass, $E_1$ corresponds to $T_e^{\rm max}=T_{e,1}$,
and $d\sigma_{\nu e}(E,T_e)/dT_e$ is the differential cross section for 
$\nu$-$e$ scattering \cite{nu-e-cross}. In Eq.~(\ref{eq:es}), the sum runs over
$F_{\nu_e}$, $F_{\bar\nu_e}$, $F_{\nu_\mu+\nu_\tau}$, and 
$F_{\bar\nu_\mu+\bar\nu_\tau}$, with the last two fluxes multiplied by
$d\sigma_{\nu_x e}/dT_e$ and $d\sigma_{\bar\nu_x e}/dT_e$, respectively.

The pre-SN ES signals mostly occur at $T_e\leq 2.5$~MeV, but solar neutrinos present a high 
background at $T_e<0.8$~MeV. Taking $T_{e,1}=0.8$~MeV and $T_{e,2}=2.5$~MeV, we obtain 
$N^{\rm IH}_{\rm ES}/N^{\rm NH}_{\rm ES}$ $\approx 1.23$ for all the stellar models considered.
This ratio is insensitive to the energy and time windows. For our adopted windows,
we find $N^{\rm NH}_{\rm ES}/N^{\rm NH}_{\rm IBD}\approx 0.91$ for all of our stellar models. 
In contrast, the above ratios along with $N^{\rm NH}_{\rm IBD}/N^{\rm IH}_{\rm IBD} \approx 3.42$
give $N^{\rm IH}_{\rm ES}/N^{\rm IH}_{\rm IBD} \approx 3.8$, which greatly exceeds
$N^{\rm NH}_{\rm ES}/N^{\rm NH}_{\rm IBD}$. This large difference in $N_{\rm ES}/N_{\rm IBD}$
between the NH and IH, along with the associated insensitivity to stellar models,
provides the basis for a model-independent determination of the $\nu$MH by combining the IBD
and ES signals.

Unlike the IBD events, which can be identified by coincidence,
ES causes single hits in the detector and suffers from high background. For our adopted
energy window of $0.8 \le T_e \le 2.5$~MeV, the dominant background at JUNO is $\beta^+$ decay
of the cosmogenic $^{11}$C, with an estimated level of $\sim 2\times 10^4$ events per day 
\cite{juno}. For comparison, the predicted number, $N_{\rm ES}$, of pre-SN ES signals from 
Betelgeuse over the last day is 117.2 (143.5), 212.9 (259.0), 380.9 (467.1), or 
479.8 (592.1) for the NH (IH) and a mass of 12, 15, 20, or $25\,M_\odot$, respectively.
Therefore, the above model-independent method to determine the $\nu$MH
is practical only when the high ES background can be suppressed.
Because $^{11}\rm C$ is mainly produced by $(\gamma, n)$ spallation following the shower 
initiated by cosmic muons, a three-fold coincidence of the muon, neutron, and $^{11}$C
decay products can be used to suppress the background \cite{c11,juno}. With this possible 
experimental improvement in mind, we calculate the maximum allowed number, $N_b^{\rm ES}$, of 
ES background events so that the model-independent method can be used to determine 
the $\nu$MH at the 95\% CL with the pre-SN signals from Betelgeuse.

We define
\begin{equation}
R \equiv\frac{N'-N^{\rm ES}_b}{N-N^{\rm IBD}_b},
\end{equation}
where $N'$ and $N$, respectively, are the observed numbers of ES and IBD events including 
the associated background. The expected number, $N^{\rm IBD}_b$, of IBD background events is 
the same as in Section~\ref{sec:ibd} and assumed to be well measured. The expected number,
$N^{\rm ES}_b$, of ES background events is to be constrained but is also 
assumed to be well measured. Similarly to the analyses in Section~\ref{sec:ibd},
$N'$ and $N$ follow the corresponding Poisson distributions. To allow for
large uncertainties in the predicted numbers of signals in view of the poorly-known stellar
model of Betelgeuse, we calculate the expected numbers, 
$\tilde N^{\rm NH(IH)}_{\rm ES}$ and $\tilde N^{\rm NH(IH)}_{\rm IBD}$, of ES and IBD signals,
respectively, for the NH (IH) as follows. We treat the predicted number, 
$N^{\rm NH}_{\rm IBD}$, of IBD signals as a parameter. For each 
predicted $N^{\rm NH}_{\rm IBD}$, we consider that the expected 
$\tilde N^{\rm NH}_{\rm IBD}$ is uniformly distributed over $[0.5,2]N^{\rm NH}_{\rm IBD}$ 
as a conservative estimate. For each $\tilde N^{\rm NH}_{\rm IBD}$, we
generate $\tilde N^{\rm IH}_{\rm IBD}$, $\tilde N^{\rm NH}_{\rm ES}$,
and $\tilde N^{\rm IH}_{\rm ES}$ by sampling Gaussian distributions for the ratios 
$\tilde N^{\rm NH}_{\rm IBD}/\tilde N^{\rm IH}_{\rm IBD}$, 
$\tilde N^{\rm NH}_{\rm ES}/\tilde N^{\rm NH}_{\rm IBD}$, and 
$\tilde N^{\rm IH}_{\rm ES}/\tilde N^{\rm NH}_{\rm ES}$. Based on our stellar models, we
adopt central values of 3.42, 0.91, and 1.23, respectively, for these distributions, with a 
common $1\sigma$ relative uncertainty of 5\% (including the $\sim 1$--2\% variations of 
the above ratios due to uncertainties in the vacuum neutrino mixing parameters \cite{pdg}).

\begin{figure}[htbp] 
\centering
\includegraphics[width=8.cm]{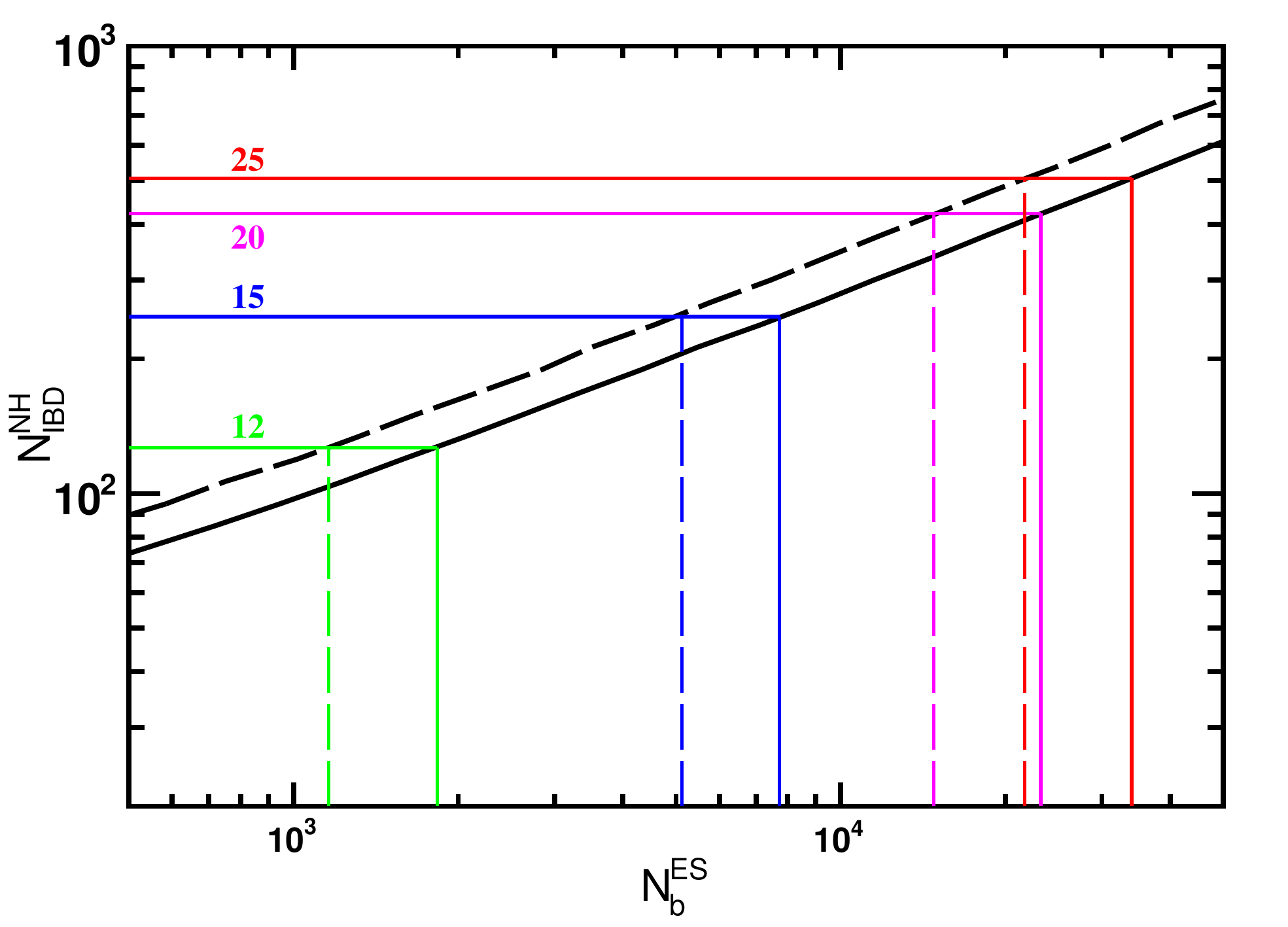}        
\caption{ 
Combinations of $N^{\rm NH}_{\rm IBD}$ and $N_b^{\rm ES}$ required to determine 
the $\nu$MH at the 95\% CL. The horizontal solid lines indicate the predicted $N_{\rm IBD}^{\rm NH}$ 
over the last day before the core collapse of Betelgeuse at $d=222$ pc for an assumed 
mass of 12, 15, 20, or $25\,M_\odot$. The solid curve ignores the pre-SN $\nu_e$
produced by weak nuclear processes, whereas the dashed curve represents an estimate of their
maximum effect.
}
\label{fig:ES_b}    
\end{figure}

For each $N^{\rm NH}_{\rm IBD}$, we generate $10^6$ sets of $N'_{\rm NH(IH)}$ 
and $N_{\rm NH(IH)}$ to calculate the distribution $P_{\rm NH(IH)}$ of $R_{\rm NH(IH)}$, 
which peaks at $R_{\rm NH(IH)}\approx N^{\rm NH(IH)}_{\rm ES}/N^{\rm NH(IH)}_{\rm IBD}\approx 0.91$
(3.8). The distributions $P_{\rm NH}$ and $P_{\rm IH}$ cross at $R_{\rm NH}=R_{\rm IH}=R_0$.
Similarly to the analyses with IBD events only, we consider that the $\nu$MH can be determined at 
the 95\% CL when
\begin{align}
\int_{-\infty}^{R_0}P_{\rm NH}dR_{\rm NH}=\int_{R_0}^\infty P_{\rm IH}dR_{\rm IH}= 0.95.
\end{align}
The combinations of $N_{\rm IBD}^{\rm NH}$ and $N_b^{\rm ES}$ corresponding to the above
criterion are shown as the solid curve in Fig.~\ref{fig:ES_b}, where the predicted values of 
$N_{\rm IBD}^{\rm NH}$ for our stellar models are also indicated. It is reasonable to assume
that our 15 and $25\,M_\odot$ models provide the limiting cases for Betelgeuse, especially 
when the results shown in Fig.~\ref{fig:ES_b} allow for 
a factor of 2 uncertainty in the model prediction. Accordingly, we conclude that the pre-SN IBD 
and ES signals from Betelgeuse over the last day can be used to determine the $\nu$MH at the 
95\% CL in a model-independent manner if the ES background in JUNO can be reduced from 
$N_b^{\rm ES}\sim 2\times 10^4$ by a factor of $\sim 2.5$. If our $12\,M_\odot$ model fits
Betelgeuse better, the reduction needs to be by a factor of $\sim 10$.

So far we have ignored the pre-SN $\nu_e$ produced by weak nuclear processes in stars.
In view of the theoretical uncertainties associated with these $\nu_e$, we estimate their 
maximum effect by treating their contribution to the ES signals as additional uncertainties 
in the ratios $N^{\rm NH}_{\rm ES}/N^{\rm NH}_{\rm IBD}$ and $N^{\rm IH}_{\rm ES}/N^{\rm NH}_{\rm ES}$.
For a generous estimate, we consider that these $\nu_e$ are up to $\sim 50\%$ of those 
produced by $e^\pm$ pair annihilation in the relevant energy window \cite{patton2}. 
As increasing $F_{\nu_e}^{(0)}$ by $\sim 50\%$ increases $N^{\rm NH}_{\rm ES}/N^{\rm NH}_{\rm IBD}$ 
and $N^{\rm IH}_{\rm ES}/N^{\rm NH}_{\rm ES}$ by $\sim 15\%$ and $\sim 8\%$, respectively, we
adopt larger $1\sigma$ relative uncertainties of 20\% and 10\% for the Gaussian distributions
of $N^{\rm NH}_{\rm ES}/N^{\rm NH}_{\rm IBD}$ and $N^{\rm IH}_{\rm ES}/N^{\rm NH}_{\rm ES}$,
respectively, and repeat the calculations described above. The results are shown
as the dashed curve in Fig.~\ref{fig:ES_b}. It can be seen that the maximum 
effect of the pre-SN $\nu_e$ produced by weak nuclear processes is to require a further 
reduction of the ES background by a factor of $\sim 1.5$ for a model-independent determination 
of the $\nu$MH with pre-SN neutrinos from Betelgeuse.

\section{Discussion and conclusions}
We have presented quantitative analyses of pre-SN neutrino signals at JUNO
as potential probes of the $\nu$MH. Using the IBD events alone, we have
considered three cases, for all of which determination of the $\nu$MH
requires accurate stellar models of pre-SN neutrino emission. In the ideal case
where the distance to the source is known exactly and the uncertainty 
in the predicted number, $N_{\rm IBD}$, of IBD events is 10\%, the $\nu$MH can be 
determined at $\gtrsim$ 95\% CL with pre-SN IBD signals over the last day from stars 
of 12, 15, 20, and $25\,M_\odot$ within $\approx 0.44$, 0.6, 0.8, and 0.88 kpc, 
respectively. In the case where the stellar model for the nearby Betelgeuse is known, 
determination at this level requires an uncertainty of $\lesssim 30\%$ in the 
predicted $N_{\rm IBD}$. In the more realistic case where our 15 and $25\,M_\odot$
models provide the limiting cases for Betelgeuse, 
this uncertainty is restricted to $\lesssim 14\%$. 
With the current measurement of $d=222^{+48}_{-34}$~pc for the distance to Betelgeuse 
\cite{distance}, the error in $d$ already gives a $\sim 30\%$ uncertainty in the
predicted $N_{\rm IBD}$. Even if this distance can be measured precisely, the
required uncertainty of $\lesssim 14$--30\% in the prediction is difficult
to achieve for stellar models.

We advocate a model-independent determination of the $\nu$MH using
both the pre-SN IBD and ES events at JUNO. This determination relies on the 
large difference in $N_{\rm ES}/N_{\rm IBD}$ between the NH and IH, as well as 
the insensitivity of this ratio to stellar models. The key issue here is the ES 
background in the adopted energy window of $0.8\leq T_e\leq 2.5$~MeV, 
which is dominated by $\beta^+$ decay of the cosmogenic $^{11}$C. 
Our analyses show that if our 15 and $25\,M_\odot$ models provide the limiting cases for 
Betelgeuse, using its pre-SN IBD and ES signals to determine the 
$\nu$MH at the $\gtrsim 95\%$ CL requires this background to be $\lesssim 8\times10^3$ 
events per day. With the background currently estimated to be $\lesssim 2\times10^4$ 
events per day, the required reduction by a factor of $\sim 2.5$ is possible by using 
coincidence of the background events \cite{c11,juno}. Even if our $12\,M_\odot$ model
fits Betelgeuse better, the required reduction by a factor of
$\sim 10$ might still be feasible. In any case, however, 
a further reduction by a factor of $\sim 1.5$ 
might be required when uncertainties associated with the pre-SN $\nu_e$ produced by weak 
nuclear processes are taken into account. On the other hand, measuring solar neutrinos
at JUNO precisely may allow us to use the ES signals with $T_e<0.8$~MeV, which would
increase the pre-SN signals significantly, thereby relaxing the requirement of
the cosmogenic background reduction.

The pre-SN $\nu_e$ of $\sim 5$--10~MeV from weak nuclear processes 
produce signals in both charged-current and neutral-current channels
at DUNE. These signals can, in principle, provide a model-independent 
determination of the $\nu$MH, which merits a quantitative assessment.
We note, however, that the relevant event rates are low and have 
significant theoretical uncertainties. 

A large number of neutrino events can be detected from a Galactic SN
(e.g., \cite{snnudet1,snnudet2,juno,horiuchi18}). Flavor evolution of
SN neutrinos, however, is complicated by details of their emission,
SN dynamics, and collective oscillations (e.g., \cite{cno,snnumo}), which may
make it difficult to determine the $\nu$MH with these neutrinos.
Therefore, pre-SN neutrinos are not only precursors to their SN counterpart, 
but also complementary probes of neutrino physics. We consider it an exciting
possibility to determine the $\nu$MH with pre-SN neutrinos from
Betelgeuse and urge that background reduction at JUNO be
explored for the model-independent determination presented here.

\section*{Acknowledgments}
This work was supported in part by the Deutsche Forschungsgemeinschaft 
(279384907-SFB 1245, GG), the US Department of Energy (DE-FG02-87ER40328, YZQ), 
the Australian Research Council (FT120100363, AH), the National Natural Science 
Foundation of China (11655002, TDLI), and the Science and Technology Commission 
of Shanghai Municipality (16DZ2260200, TDLI).


\begin{thebibliography}{000}

\bibitem{Itoh96}
N. Itoh, H. Hayashi, A. Nishikawa, and Y. Kohyama, Astrophys. J. Suppl. Ser. 102, 411 (1996).

\bibitem{guo}
G. Guo and Y.-Z. Qian, Phys. Rev. D 94, 043005 (2016). 

\bibitem{juno} 
F. An et al. (JUNO), J. Phys. G 43, 030401 (2016).

\bibitem{dune}
J. Strait et al., arXiv:1601.05823. 

\bibitem{odrzywolek04}  
A. Odrzywolek, M. Misiaszek, and M. Kutschera, Astropart.Phys. 21, 303 (2004).

\bibitem{odrzywolek10} 
A. Odrzywolek and A. Heger, Proceedings, 16th Cracow Epiphany Conference on Physics in Underground Laboratories and its Connection with LHC, Acta Phys. Polon. B 41, 1611 (2010).

\bibitem{kato15} 
C. Kato, M. D. Azari, S. Yamada, K. Takahashi, H. Umeda, T. Yoshida, and K. Ishidoshiro,
Astrophys. J. 808, 168 (2015). 

\bibitem{kamland16}
K. Asakura et al. (KamLAND), Astrophys. J. 818, 91 (2016).

\bibitem{yoshida16}
T. Yoshida, K. Takahashi, H. Umeda, and K. Ishidoshiro, Phys. Rev. D 93, 123012 (2016).

\bibitem{kato17}  
C. Kato et al., Astrophys. J. 848, 48 (2017).    
  
\bibitem{patton1}
K. M. Patton, C. Lunardini, and R. J. Farmer, Astrophys. J. 840, 2 (2017). 

\bibitem{patton2}   
K. M. Patton, C. Lunardini, R. J. Farmer, F. X. Timmes, Astrophys. J. 851, 6 (2017). 

\bibitem{progenitor}
M. M. Dolan, G. J. Mathews, D. D. Lam, N. Q. Lan, G. J. Herczeg, and D. S. P. Dearborn, Astrophys. J. 819, 7 (2016).

\bibitem{distance}
G. M. Harper et al., Astrophys. J. 154, 11 (2017).  

\bibitem{msw}
L. Wolfenstein, Phys. Rev. D {\bf 17}, 2369 (1978);
S. P. Mikheyev and A. Y. Smirnov, Sov. J. Nucl. Phys. {\bf 42}, 913 (1985).

\bibitem{starosc}
A. S. Dighe and A. Y. Smirnov, Phys. Rev. D 62, 033007 (2000).

\bibitem{alex}
S. E. Woosley, A. Heger, T. A. Weaver, Rev. Mod. Phys., 74, 4 (2002).

\bibitem{numix}
M. C. Gonzalez-Garcia, M. Maltoni, and T. Schwetz, J. High Energy Phys. 11 (2014) 052.

\bibitem{pdg}
M. Tanabashi et al. (Particle Data Group), Phys. Rev. D 98, 030001 (2018) 

\bibitem{geo}
V. Strati, M. Baldoncini, I. Callegari, F. Mantovani, W.F. McDonough, B. Ricci and G. Xhixha, Prog. in Earth and Planet. Sci. 2, 5 (2015).

\bibitem{nu-e-cross}
W. J. Marciano and Z. Parsa, J. Phys. G 29, 2629 (2003). 

\bibitem{c11}
C. Galbiati, A. Pocar, D. Franco, A. Ianni, L. Cadonati, and S. Sch\"onert,
Phys. Rev. C 71, 055805 (2005).

\bibitem{snnudet1} 
K. Scholberg, Annu. Rev. Nucl. Part. Sci. {\bf 62}, 81 (2012).

\bibitem{snnudet2}
A. Mirizzi et al., Riv. Nuovo Cim. 39, no. 1-2, 1 (2016).

\bibitem{horiuchi18}
S. Horiuchi, J. P. Kneller, J. Phys. G 45, 043002 (2018).

\bibitem{cno}
H. Duan, G. M. Fuller, and Y.-Z. Qian, 
Annu. Rev. Nucl. Part. Sci. {\bf 60}, 569 (2010).

\bibitem{snnumo}
K. Scholberg, J. Phys. G45, 014002 (2018).  





 
\end{thebibliography}
\end{document}